\documentclass[aps,twocolumn,amsfonts,showpacs]{revtex4}
\usepackage{epsfig,amsopn,amsmath}
\usepackage{multirow,bigdelim}

\usepackage{color}
\usepackage{lipsum}

\begin{document}

\title{The influence of the composition of tradeoffs on the generation of differentiated cells}

\author{Andr\'e Amado}
\affiliation{%
Departamento de F\'{\i}sica, Universidade Federal de Pernambuco, \\
52171-900 Recife-PE, Brazil
}%

\author{Paulo R. A. Campos}\email{prac@df.ufpe.br}
\affiliation{%
Departamento de F\'{\i}sica, Universidade Federal de Pernambuco, \\
52171-900 Recife-PE, Brazil
}%

\begin{abstract}
We study the emergence of cell differentiation under the assumption of the existence
of a given number of tradeoffs between genes encoding different functions.  In the model the viability of colonies is
determined by the capability of their lower level units to perform different functions, which is implicitly determined
by external chemical stimuli. Due to the existence of tradeoffs it can be evolutionarily advantageous to evolve the division of labor
whereby the cells can suppress their contributions to some of the activities through 
the activation of regulatory genes, which in its turn
inflicts a cost in terms of fitness. Our simulation results show that cell differentiation is more likely as the number of tradeoffs
is increased but the outcome also depends on their strength. We observe the existence of critical values for 
the minimum number of tradeoffs and their strength beyond that maximum cell differentiation can be attained.  Remarkably, we observe
the occurrence of a maximum tradeoff strength beyond that the population is no longer viable imposing an upper tolerable level
of constraint at the system. This tolerance is reduced as the number of tradeoffs grows.
\end{abstract}


\maketitle

\section{Introduction}\label{section}
The evolution of tradeoffs is an issue of intense debate in the current literature as 
tradeoffs have played a central role in shaping life histories and ecological/evolutionary
dynamics in nature \cite{SaekiOikos2014}. The tradeoff theory is key for our progress in evolutionary understanding.
Traits are often linked in ways that prevent simultaneous optimization of all of them, as they reflect biophysical
compromises \cite{GudeljJEVOLBIOL2007,MeyerNature2015}.
Tradeoffs are referred to as the cost paid in terms of fitness when a beneficial change in one trait
is linked to a detrimental change in another \cite{StearnsFuncEcol1989}.
With the massive
available data from experiments, especially in microbial populations,
this problem has been more effectively
addressed \cite{MacLeanHeredity2008}. Today it is also known that shapes and magnitudes of tradeoff relationships are strongly influenced
by the environment \cite{JessupEcolLett2008}.

\bigskip

The presence of tradeoffs is one of the main conditions that are fulfilled by most biological systems
for the appearance of the division of labor \cite{RuefflerPNAS2012}, being its emergence concomitantly favored 
 by other factors \cite{GavriletsPlosCompBiol2010,RuefflerPNAS2012,BozaPlosCompBiol2014}.
One of these factors favoring the emergence of division of labor upon the existence of tradeoffs
is developmental plasticity  \cite{GavriletsPlosCompBiol2010}. 
 Developmental plasticity refers to the genotype's ability to change its developmental processes
and phenotypic outcomes in response to environmental changes, and influences the trait expression. 
Today it is well established that developmental plasticity is critical to the promotion of evolutionary 
innovation \cite{MoczekProcRSocB2011}. Indeed, plastic responses to environmental variation plays
a key role for species to develop \cite{HoltTREE1990,HoffmannNature2011}. 

\bigskip

The main goal of the present study is to address the emergence of labor division through the differentiation of cells initiating 
from undifferentiated units. A key feature of the modelling is the use of the well-grounded acquaintance 
that formerly diversity in multicellular organisms
stems from changes in the regulatory interactions that drive gene expression \cite{NadalNatureGen2011}, and so many of the
requirements for multicellularity evolved in unicellular ancestors \cite{GrosbergAnnRevEcol2007}. 
Sophisticated sensing mechanisms and signal transduction systems in Eukariotic cells allow 
accurate dynamic outcomes in response to changing environment conditions \cite{NadalNatureGen2011}.

\bigskip

\begin{figure*}
\centering
\begin{tabular}{cc}
	\includegraphics[width=0.45\textwidth]{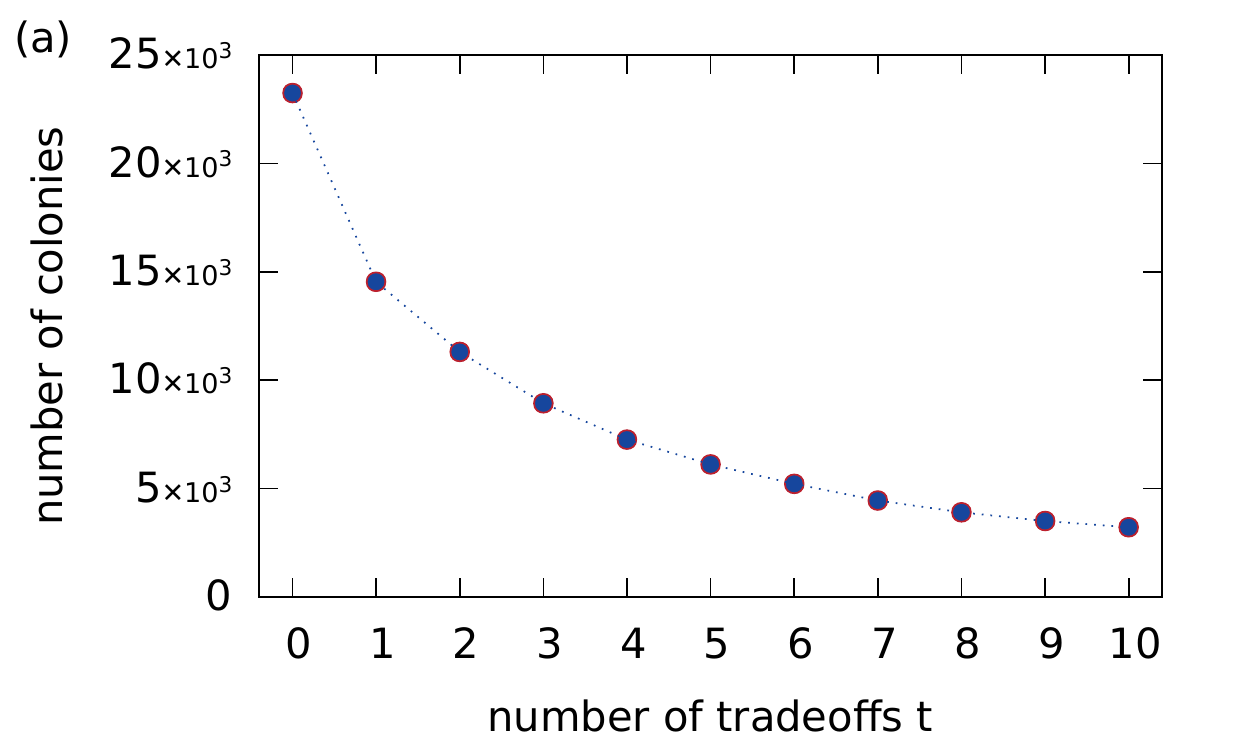}&
	\includegraphics[width=0.45\textwidth]{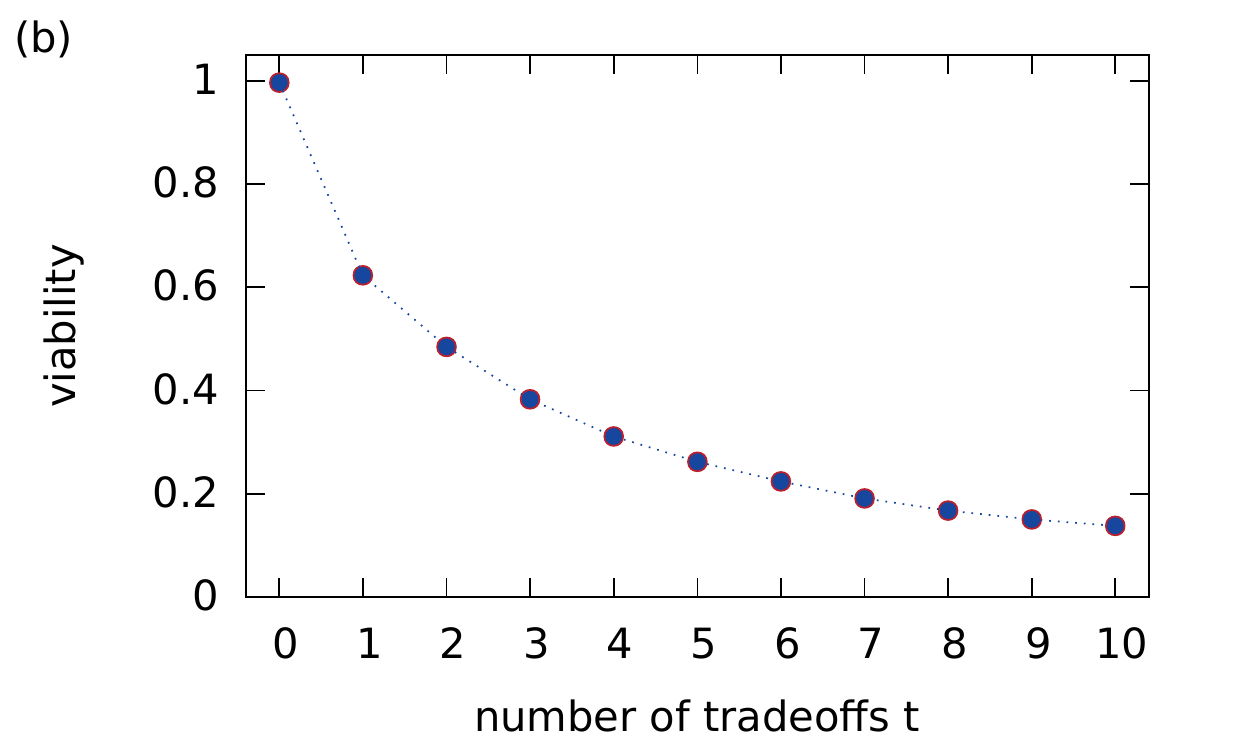}
\end{tabular}
\caption{Panel a: Average number of colonies versus the number of trade-offs $t$ for four biological functions (the remaining parameters are $s=2$, $S=16$, $\mu=10^{-5}$ and $K = 50000$).
Panel b: Average viability of a colony versus the number of trade-offs $t$ for four biological functions (the remaining parameters are $s=2$, $S=16$, $\mu=10^{-5}$ and $K = 50000$).
Each point is an average over 1000 independent configurations.}
\label{fig:fig1}
\end{figure*}

Recent theoretical contributions address the evolution of multicellularity and the further specialization
of cell types, mainly focusing on the differentiation between somatic and germinative
functions \cite{MichodPNAS2007,GavriletsPlosCompBiol2010}. Here the division of labor and 
subsequent differentiation is addressed to the competition among distinct
somatic functions in a population which comprises genetically identical cells. Conditions upon the tradeoffs 
that allow cellular specialization to evolve are investigated. Cellular differentiation can evolve
as an outcome of the selective advantage brought by the division of labor between the cells. The process is entirely grounded on
 regulation mechanisms, as it influences the viability of the aggregates which are formed. 
We assume the existence of tradeoffs among a set of somatic functions. 
An illustrative instance is provided by the cyanobacteria whose cells can specialize either
in the carbon or nitrogen fixation processing \cite{ShapiroSciAmer1988,FloresNatRevMic2010}.
The underlying concept is that there is a set of essential biological functions which contributes to
the viability of the organism. These functions are subjected to biophysical constraints that
impose tradeoffs  between the different functions. These relations can lead to
specialization of cells, as it becomes advantageous to develop compartmentalized function in order 
to reduce the cost brought about by the tradeoffs. The biophysical constraints are not 
explicitly considered, as it is not our aim to propose any mechanistic model for cells, but instead our system
assumes a certain amount of randomly assigned tradeoff relationships between the different 
functions. Developmental plasticity is allowed through the presence of regulatory genes that, if activated, 
can attenuate or even completely suppress the action of specific genes whenever the cell is subordinated to a given chemical signalling.

\bigskip

The paper is organized as follows. In Section II the model is described. Section III presents 
the simulation results, and finally Section IV presents our concluding remarks.

\section{The Model}\label{section}
The population consists of asexual haploid cells. The model assumes clonal development from a unicellular spore/zygote, thus giving
rise to multicellular organisms (colonies) as cells undergo binary fission \cite{GrosbergAnnRevEcol2007,GavriletsPlosCompBiol2010}.
Under clonal development genetic variation among cell lineages is relatively low within an organism and basically 
stems from somatic mutations.
During clonal development the cell passes through binary fission processes until the colony  
at this expansion stage reaches size $S$. After 
reaching size $S$ each colony goes through an unicellular stage (propagule formation) in case
it survives viability selection. Each cell can give rise to new daughter colonies with a given probability that depends on its fertility $f$.

\bigskip

\noindent
The tradeoffs are represented by a matrix $\mathbf{T}=\{T_{ij}\}$ that measures the strength
of the ascendency of the somatic function $j$ on the function $i$. The main gene $Y_i$ encodes for 
function $i$, whereas the regulatory genes $y_{ik}$ regulates the 
expression of gene $Y_i$ in a cell undertaking activity $k$. As we know, a cell is a complex self-regulated system
that responds in different ways to different set of chemical signalling \cite{IspolatovProcRSocB2011}. Here we assume that 
there is temporal segregation of 
incompatible activities  of cells which are induced by these external chemical signals. 
The activation of regulatory genes brings a cost, estimated as $c(y_{ik})$. It is assumed that  $c(y_{ik})$ 
is a decreasing function of $y_{ik}$. As the system evolves
incompatible cellular processes tend to suppress the expression of genes encoding other functions \cite{TamDevelopment1991}, 
thus contributing to the formation of aggregates with permanently specialized cellular functions.

\bigskip


\begin{figure*}
\centering
\begin{tabular}{cc}
	\includegraphics[width=0.45\textwidth]{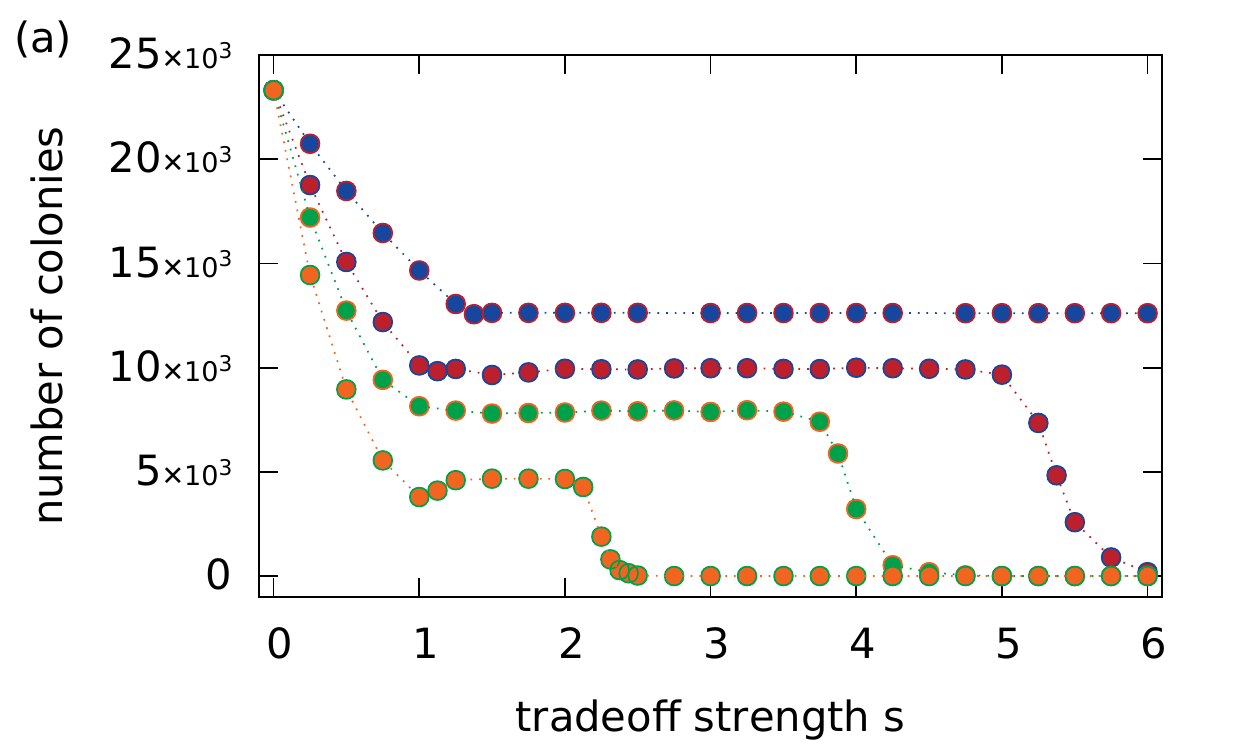}&
	\includegraphics[width=0.45\textwidth]{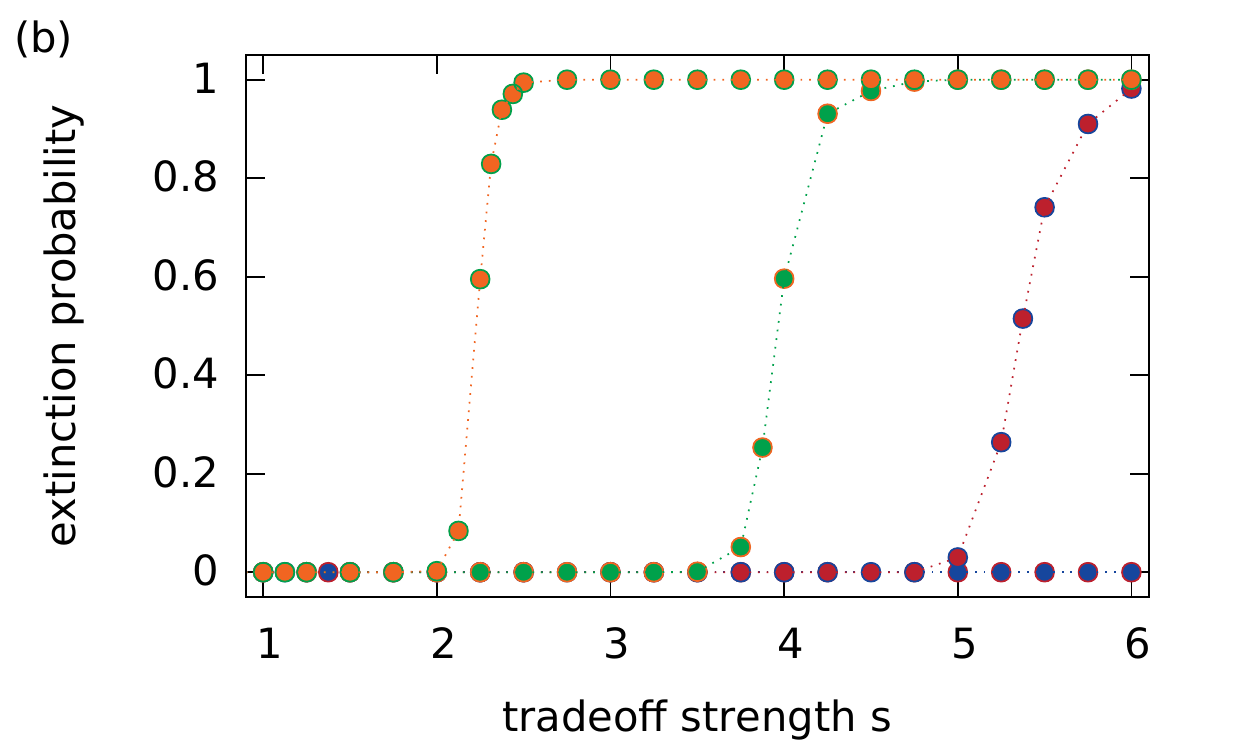}
\end{tabular}
\caption{Panel a: Average number of colonies versus the strength of the trade-off $s$ for three biological functions and one (dark blue), two (red), three (green) and six (orange) trade-offs (the remaining parameters are $S=16$, $\mu=10^{-5}$ and $K = 50000$). 
Panel b: Extinction probability versus the strength of the trade-off $s$ for three biological functions and one (dark blue), two (red), three (green) and six (orange) trade-offs (the remaining parameters are $S=16$, $\mu=10^{-5}$ and $K = 50000$). 
Each point is an average over 1000 independent configurations.}
\label{fig:fig2}
\end{figure*}

\noindent
The contribution of a cell subject to chemical signal $k$ to the overall somatic function
$i$ ($f_{ik}$) of the colony is then calculated as
\begin{align}\label{regulation}
f_{ik} = \left[(1-y_{ik})Y_i\right]^{T_{ii}} c(y_{ik}) \prod_{j\neq i}\left[1-(1-y_{jk})Y_j\right]^{T_{ij}} c(y_{jk}).
\end{align}
We see that the direct effects of the major genes $Y_{i}$ increase the corresponding fitness components, whereas
 there are indirect negative effects of the other genes $Y_{j}$, $i \neq j$, on the major gene, and thereby Eq. (\ref{regulation}) captures
 the essence of the tradeoff relationships. The 
case $T_{ij}=0$ reflects the non-existence of tradeoff between the corresponding pair of genes. 
As the mechanisms of gene suppression and developmental plasticicy embody a cost in fitness terms, it is incorporated
into the estimation of $f_{ik}$ through the cost function  $c(y_{ik})$, which is a decreasing function of the regulation effect 
$y_{jk}$. This means that the stronger suppresion is more costly it becomes  \cite{GavriletsPlosCompBiol2010,VanTienderenEvolution1991}. 
Without this regulation, Eq. (\ref{regulation}) reduces to its simplest form
\begin{align}
	f_i = Y_i^{T_{ii}}\prod_{j\neq i}\left[1-Y_j\right]^{T_{ij}}.
\end{align}
The cost function  $c(y)$ is given by a Gaussian function $c(y)=\exp(-\frac{1}{2}\frac{y^2}{\sigma_{y}^{2}})$.  The full 
expression (Eq. (\ref{regulation})) can be compactly written as 
\begin{align}
f_{ik} = \prod_j \left|1-\delta_{ij}-(1-y_{jk})Y_j\right|^{T_{ij}} c(y_{jk}).
\end{align}

\noindent 
To estimate the viability of a given colony, first the average  
contribution of the cells to the fitness components of the group is computed \cite{MichodPNAS2007}, i.e., 
\begin{align}
	f_i = \frac{1}{N_c}\sum_k f_{ik}
\end{align}
where $N_c$ stands for the number of cells. The viability is then calculated as 
the geometric mean of the $f_{i}$-values, i.e., all functions are considered to be essential. 
Therefore the viability is expressed as 
\begin{align}
	v = \sqrt[\leftroot{-2}\uproot{5}N_f]{\ \prod_i f_i\,}
\end{align}
and $N_f$ corresponds to the number of distinct biological functions. In Ref. \cite{GavriletsPlosCompBiol2010}
the viability of the colony is completely determined by a single somatic function. From 
the measurement of the viability  it ensues  that the likelihood a given organisms survives  
to reproduction age equals \cite{BevertonBook2012}
\begin{align}\label{probability}
	\left[1+(S-1)\frac{N}{K v}\right]^{-1}.
\end{align}
The above equation is a modified version of the  Beverton-Holt stock-recruitmen model which assumes that the per capita number of offspring
is inversely proportional to a linearly increasing function of the number of mature colonies \cite{BevertonBook2012}. 
In Eq. (\ref{probability}) $N$ corresponds to the number of 
colonies and $K$ denotes the maximum carrying 
capacity of the population. As aforementioned, $S$ is the size of colony just before the unicellular stage takes place.

\subsection{Summary of the parameters of the model}
\begin{description}
\item [$s$ (tradeoff strength):] In the simplest case  the tradeoff strength is uniform, $s$, over all the tradeoffs. 
Therefore, under the assumption of an uniform tradeoff strength $T_{ij}$ is either equal to zero (if there is no tradeoff between
a given pair of genes) or $s$. The assumption of an uniform tradeoff strength will be released later. In such situation, the strength $s$
is not a constant but rather taken from a given probability distribution.   
\item [$f$ (fertility):] After surviving viability selection each cell of the colony can give rise to a newly formed colony 
with probability $f$, the fertility of the cell.
\item [$\mu$ (mutation probability):] During cell division there exists an uniform probability of mutation per gene, $\mu$. If a mutation
takes place in a given gene $j$, $Y_{j}$  (In case it is a major gene) or $y_{jk}$ (in case it is a regulatory gene) changes to a randomly chosen value from an uniform distribution $[0,1)$. 
\item [$K$ (maximum carrying capacity):] The maximum carrying capacity, $K$, corresponds to the population size
upon maximum fertility, $f=1$ (all cells can successfully establish a new colony) and maximum viability, $v=1$.
\item [$t$ (number of tradeoffs):] If the number of biological activities is $N_f$ there are up to $N_f(N_f-1)$ tradeoffs 
(that corresponds to the number of degrees of freedom of the $T_{ij}$ matrix), so $t \leq N_f(N_f-1)$.
\end{description}

\section{Results}\label{section}
As aforementioned, the strength of tradeoffs between a pair of genes is better described through  the tradeoff matrix $T_{ij}$
\begin{equation}
\mathbf{T}=
  \begin{bmatrix}
    T_{11} & T_{12} & \cdots & T_{1M} \\
    T_{21} & T_{22} & \cdots & T_{2M} \\
    \vdots & \vdots & \ddots & \vdots \\
    T_{M1} & T_{M2} & \cdots & T_{MM} 
  \end{bmatrix}
\end{equation}
where the non-diagonal elements are randomly ascribed. The number
of non-null off-diagonal elements is $t$, and the strength of the tradeoff between a given pair of genes is either assumed to be
constant $T_{ij}=s$ or taken from a given probability distribution. Simulation results for the two cases are presented separately.

\begin{figure*}
\centering
\begin{tabular}{cc}
	\includegraphics[width=0.45\textwidth]{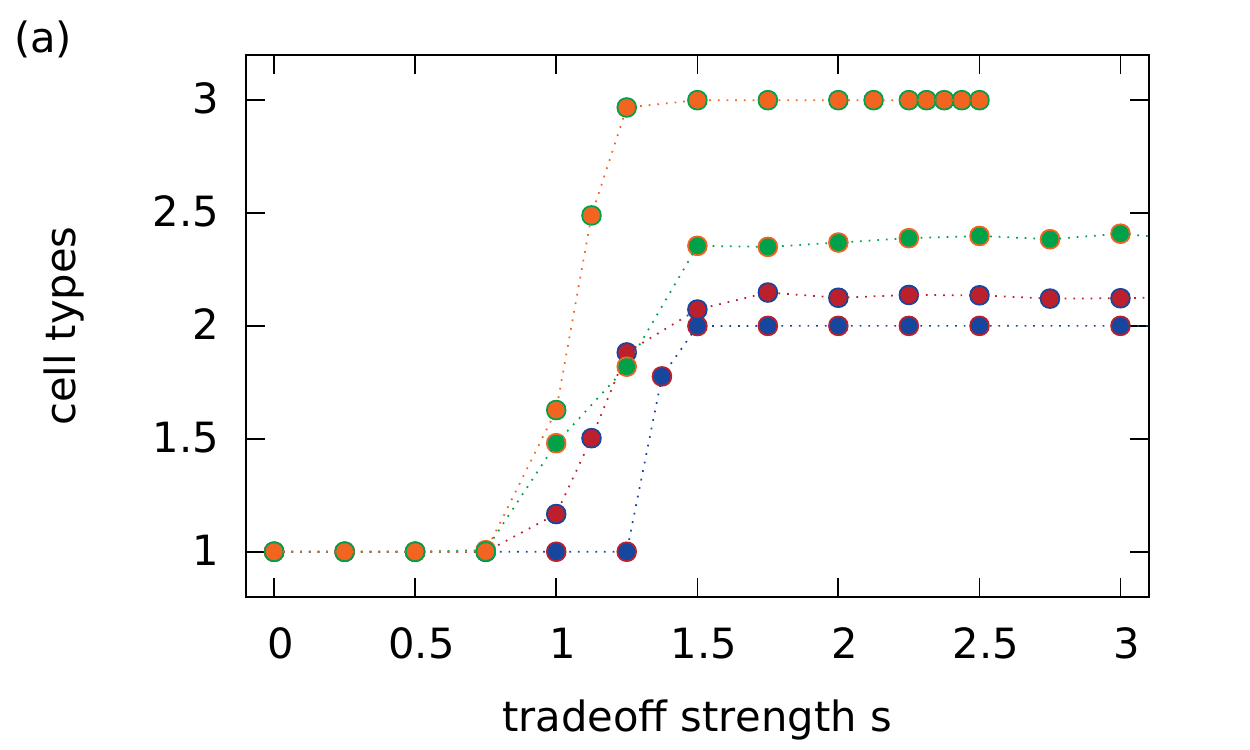}&
	\includegraphics[width=0.45\textwidth]{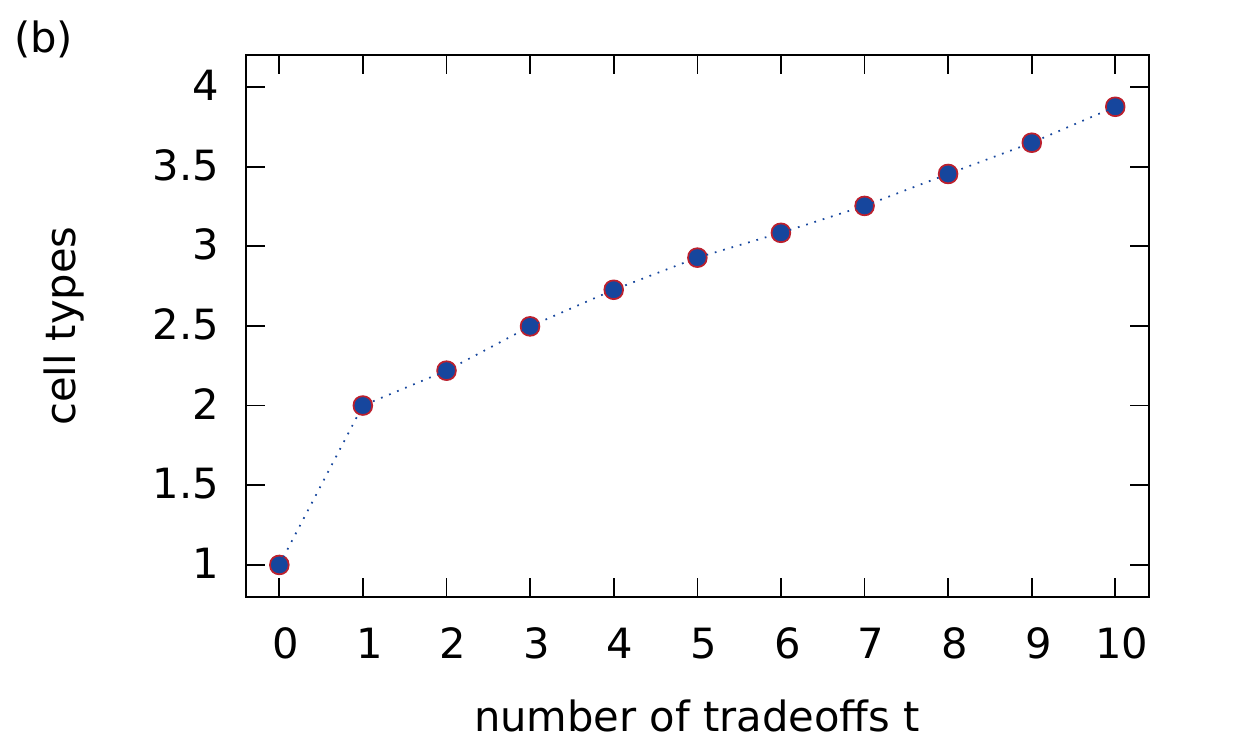}
\end{tabular}
\caption{Panel a: Average number of cell types versus the strength of the trade-off $s$ for three biological functions and one (dark blue), two (red), three (green) and six (orange) trade-offs (the remaining parameters are $S=16$, $\mu=10^{-5}$ and $K = 50000$). 
Panel b: Average number of cell types versus the number of trade-offs $t$ for four biological functions (the remaining parameters are $s=2$, $S=16$, $\mu=10^{-5}$ and $K = 50000$).
Each point is an average over 1000 independent configurations.}
\label{fig:fig3}
\end{figure*}

\subsection{Constant tradeoff strength}

Simulations were run for different number of somatic functions. Unless stated otherwise it is assumed that 
number of cell before the unicellular stage is $S=16$, mutation probability  $\mu=10^{-5}$, fertility $f=0.5$, tradeoff strength 
$s=2$ and carrying capacity $K=50,000$. 

\noindent Figure \ref{fig:fig1}a shows that the number of colonies
is a decreasing function of the number of tradeoffs. As the number of tradeoffs is augmented more specialization is required
in order to keep the colony functional. At the individual level, specialization brings a cost in terms of fitness, though it
provides a benefit at the group level. This condition is better understood if one looks at the dependence of the mean viability
on the number of tradeoffs $t$ (see Fig. \ref{fig:fig1}b). As can be noticed the viability is a monotonic
decreasing function of the number of trade-offs. If there are no tradeoffs the mean viability goes to one, meaning that all traits
can be maximised simultaneously as there are no constraints. As tradeoffs are added specialization  requires the suppression of 
the expression of more genes entailing a greater cost in terms of fitness.

\bigskip 
In the following, Figure \ref{fig:fig2}a explores the effect of the tradeoff strength on the evolution of the system. 
In the plot the number of colonies is shown as a function of the tradeoff strength, $s$, for several values
of the number of tradeoffs $t$. As a general scenario, it is verified that the
augmentation of the strength of the interaction between the genes produces detrimental effect on the viability,  and thereby 
reducing the
number of colonies at the stationary state. From the plot it is possible to remark another interesting pattern. As the tradeoff strength
is enlarged up to a certain value of $s$ the number of colonies shrinks, beyond that point the
number of colonies remains roughly constant. Subsequently, 
there exists a second critical value of the tradeoff strength at which
 the population is no longer viable and then the population 
goes extinct (number of colonies goes to zero). This is also corroborated from Fig. \ref{fig:fig2}b that exhibits
the probability of population extinction versus $s$. One notices the occurrence of a clear transition from a 
regime in which the population always persists (probability of extinction equals zero) to a regime where the population 
is always doomed to extinction and is no longer viable (probability of extinction equals one). 
As the number of tradeoffs increases the transition region becomes sharper. 
There are two important features worth mentioning. First, the critical tradeoff strength at which the population 
is no longer viable decreases with the number of tradeoffs, i.e., the transition region is shifted
towards lower $s$, and thereby the tolerance to the strength of the tradeoff decreases 
with the number of tradeoffs. Second, the range of the tradeoff strength
at which the number of colonies remains constant is also shortened with the number of tradeoffs. 

\begin{figure*}
\centering
\begin{tabular}{cc}
	\includegraphics[width=0.45\textwidth]{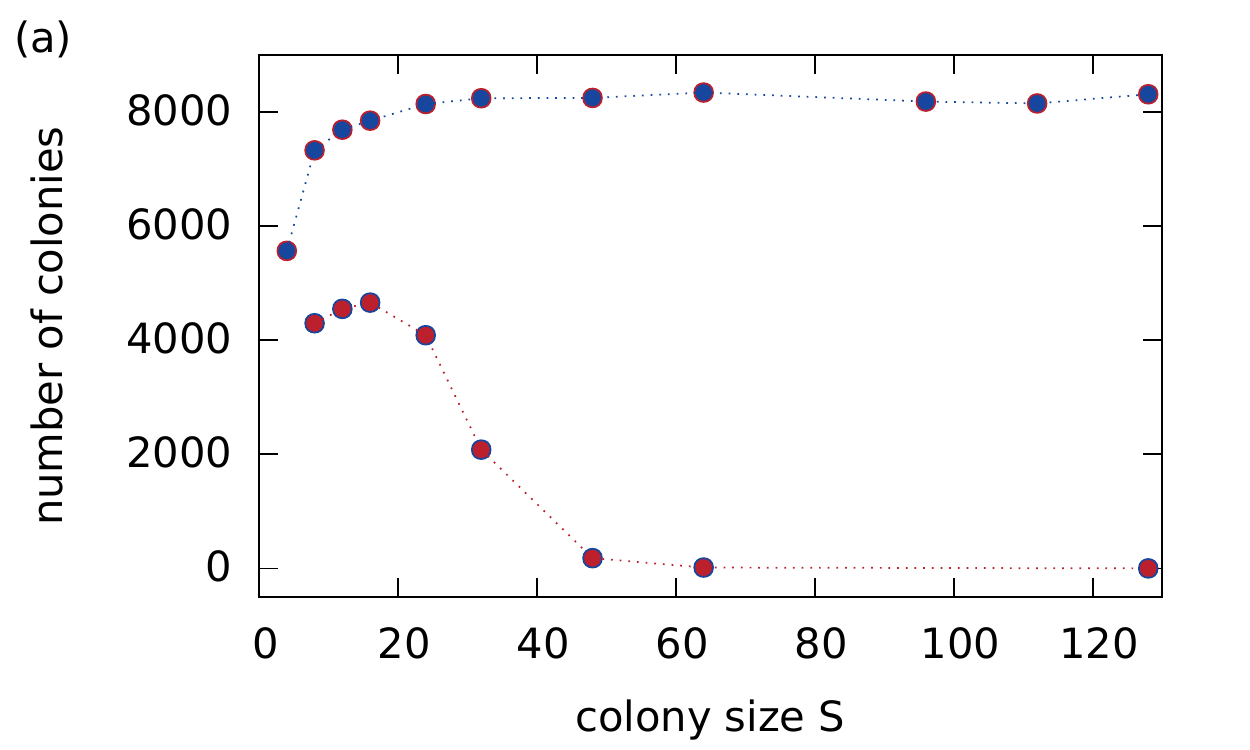}&
	\includegraphics[width=0.45\textwidth]{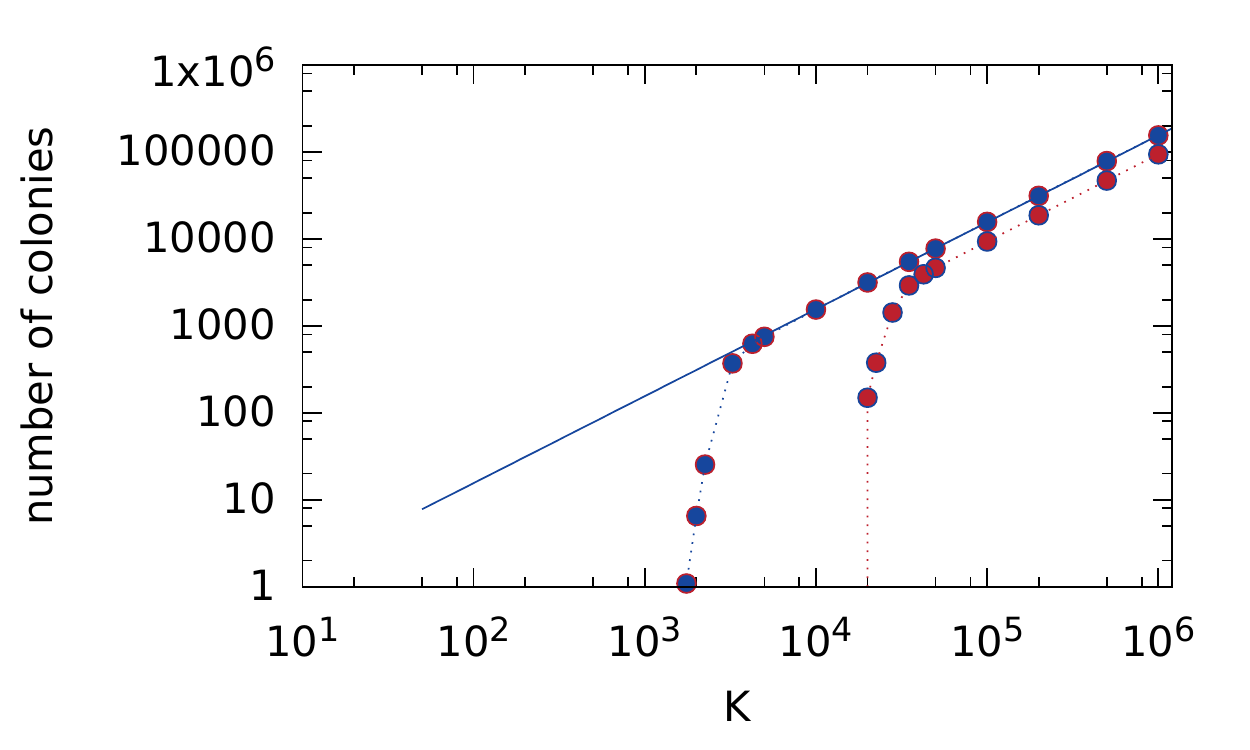}
\end{tabular}
\caption{Panel a: Average number of colonies against its size just before the reproduction stage $(S)$, for three 
biological functions and three trade-offs (the remaining parameters are $s=2$, $\mu=10^{-5}$ and $K = 50000$).
Panel b: Average number of colonies versus the maximum carrying capacity of the system ($K$) for three biological functions and three (blue) and six (red) trade-offs (the remaining parameters are $s=2$, $S=16$ and $\mu=10^{-5}$). The straight line is a linear fit of slope $+1$ as expected in the limit of large $K$ (please see Eq. (\ref{probability})). 
Each point is an average over 1000 independent configurations. }
\label{fig:fig4}
\end{figure*}

As a next step, now we will see how the evolution of the population subject to tradeoffs can ultimately drive its constituents 
to the process of differentiation. As a criterion for establishment of differentiation among the cells we propose as a metric the
distance $d$ between the response to two stimuli $i$ and $j$, which is calculated as 
\begin{align}
d_{ij} = \sqrt{\frac{\sum_{k=1}^{n_f} (y_{ki}-y_{kj})^2}{N_f}},
\end{align}
where $N_{f}$ denotes the number of biological functions. If the distance $d$ is higher than $d_{c}$ ($d_{ij} > d_{c}$) one
considers that there is a differentiated response to those stimuli (i.e. cells subject to different stimuli 
differentiate into different types). Since $y_{ij}\in[0,1]$, $d_{ij}$ also lies in the range between $0$ and $1$. 
For our purposes the threshold $d_{c}$ is set at $d_{c}=0.2$, which demonstrated to provide 
a good criterion for determining the differentiation among the cells. 
The number of differentiated types is then the number of different 
responses to the different stimuli, that is, the number of different phenotypes found in 
the cells of the same organism. The number of differentiated cells (cell types) are 
presented when the number of functions is equal to three (see Figure \ref{fig:fig3}). Therefore, in this case 
the maximum number of distinct cell types that can be reached is also equal to three.  
Figure \ref{fig:fig3}a 
shows the ultimate number of cell types versus the tradeoff strength $s$ and distinct values of the number of tradeoffs $t$. 
It is pretty clear that the existence of tradeoffs can strongly favor the emergence of cell differentiation. This can be achieved
by either increasing the number of tradeoffs, i.e. the number of non-null elements of the tradeoff matrix $T_{ij}$, or increasing the
strength of those tradeoffs. As the number of functions rises it ensues that the minimum number of tradeoffs needed 
to attain maximum differentiation is also augmented. It also follows that the lesser the number of tradeoffs the larger 
the tradeoff strength must be in order to produce maximum differentiation. In Figure \ref{fig:fig3}b the 
tradeoff strength is now held at
$s=2$ and the number of tradeoffs is varied. In that case the number of biological functions is four. 
In agreement with the previous results, as number of tradeoffs $t$ grows
cell differentiation is facilitated. We see that maximum differentiation is only achieved at $t=10$.

For the sake of completeness we also survey the 
dependence of the number of colonies on the size of group just before the unicellular state $S$ 
and the carrying capacity $K$. Both quantities influence 
the survivorship, as inferred from Eq. (\ref{probability}). For number of tradeoffs equal to three the number of colonies
exhibits an abrupt growth with $S$ in the regime of small $S$ and then saturates for intermediate and large  $S$. However, if
the number of tradeoffs is enlarged ($t=6$ in the plot) we already observe an abrupt drop of the number of colonies at intermediate,
which owes to the extinction of the population. Indeed, the fall in the number of colonies is also found for $t=3$ but this effect
occurs at much larger $S$. This outcome also shows that the colony size $S$ can not be enlarged without bound as its augmentation
reduces the probability of survival. This critical colony size $S$ depends
on the number of tradeoffs $t$. 
On the other hand an increased carrying capacity allows
the population to hold a larger number of colonies, as expected (please see Fig. \ref{fig:fig4}b). Nevertheless, one can remark
that existence of minimum levels of carrying capacity $K$ in such way the population can be sustainable. The minimum value of $K$ required
to sustain the population increases with the number of tradeoffs.

\bigskip

\subsection{Variable tradeoff strength} 
\begin{figure}
\centering
\begin{tabular}{c}
	\includegraphics[width=0.45\textwidth]{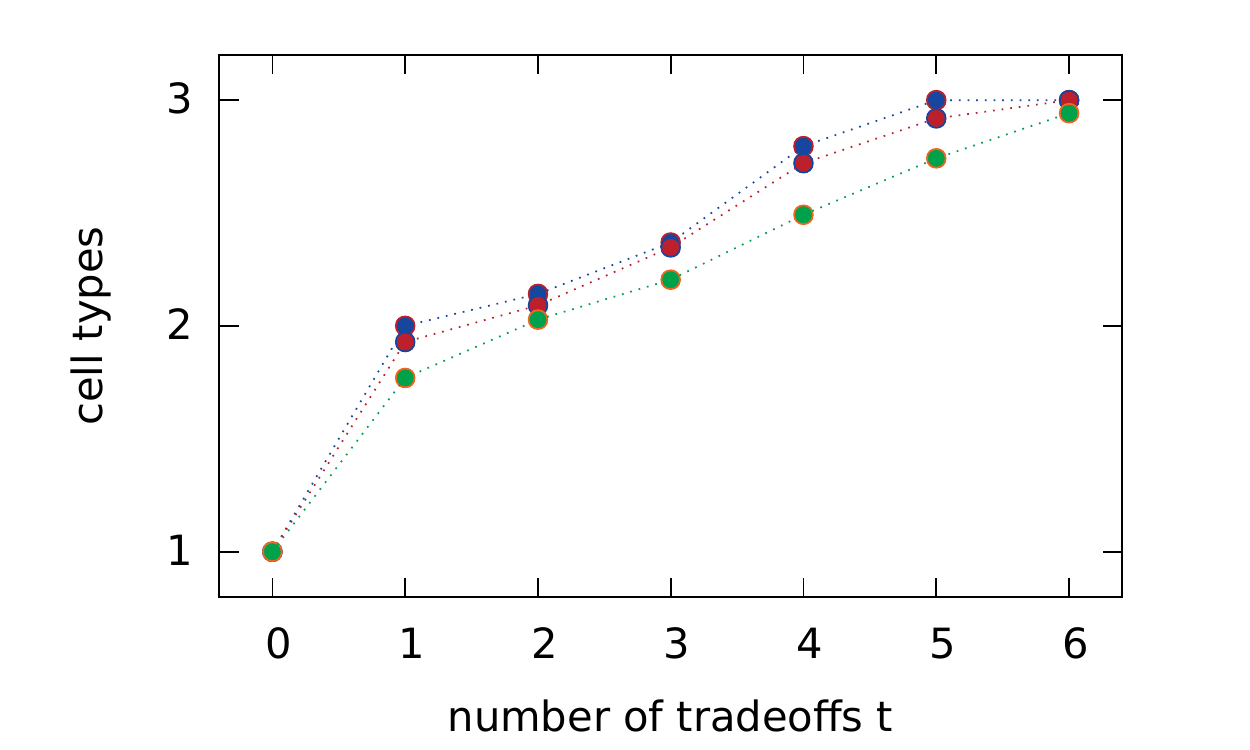}
\end{tabular}
\caption{Average number of cell types as a function of the number of tradeoffs $t$ for three biological functions. Here 
the tradeoff strength is variable and
drawn from an uniform distribution. The blue points correspond to a constant tradeoff $s=2$, the red points denote an uniform
distribution with $s \in [1,3]$, whereas the green points also denote an uniform distribution with $s \in [0.5,3.5]$. 
The remaining parameters are $S=16$, $\mu=10^{-5}$ and $K = 50000$. Each point is an average over 1000 independent configurations.}
\label{fig:fig5}
\end{figure}

Here we release the assumption of constant tradeoff strength. The tradeoff strength $s$ is
a variable quantity drawn from an uniform distribution $s \in [s_{inf},s_{sup}]$, that also varies across the
different pairs of genes. Though $t$ still tunes the number of non-null off-diagonal elements. By changing $s_{inf}$ and $s_{sup}$
the mean value of $s$ and also its variance are modified. In order to compare with the outcomes of the previous section
we changed $s_{inf}$ and $s_{sup}$ such that the mean value of $s$ is kept at two, i.e. $\langle s \rangle=2$.  And so here 
we explore the effect of increasing variance on the ultimate number of differentiated cells. From Figure 
\ref{fig:fig5} one can infer that as distribution becomes broader, and so also covering 
smaller values of $s$, the number of differentiated cells becomes considerably smaller in comparison 
to the case of constant tradeoffs. At the extent the variance is reduced the number of differentiated cells readily
approaches the outcome seen for constant strength, corroborating the finding that not only the existence of tradeoffs
 but also their magnitudes are
essential for promoting cell differentiation. 

\bigskip

\section*{Conclusions}
We have investigated how cell differentiation can arise as a consequence of division of labor, which in its turn evolves due to
developmental plasticity. The study is performed
under different scenarios for the distribution of tradeoffs. Here we have assumed that fertility is constant thus
restricting the analysis to the differentiation of cells concerning their somatic functions. In the beginning of the
evolutionary process the cells are completely undifferentiated, which means that they undertake
any function regardless of the chemical stimuli. As evolution proceeds they can suppress their contributions to some of the functions and
mostly contribute to one or few tasks through the activation of regulatory genes that can suppress some of their activities
when exposed to a given chemical stimulus. Although beneficial from the group perspective, the suppression mechanism produces a cost
at the individual level. As we can tune the number
of tradeoffs but also their strength it is possible to decouple these effects on the process of cell differentiation.

Importantly, we have noticed that the tradeoffs affect not only the outcome of the division of labor but also the viability
of the population as a whole. At the same time, an increased number of tradeoffs and their strength contribute to
the development of division of labor it also reduces the average viability of the population, and in extreme scenarios can even lead
to the population extinction. The magnitude of the tradeoffs that can be tolerated 
by the population decreases with the number of tradeoffs. Although 
tradeoffs can strongly influence populaiton's viability it also enhances the likelihood of differentiation. We have observed that
maximum differentiation, when the number of cell types equals the number of functions, is reached when the number of 
tradeoffs increases, while the strength of tradeoff required to attain this outcome is reduced.

\section*{Acknowledgments}
PRAC is partially supported by Conselho Nacional de Desenvolvimento Cient\'{\i}fico e Tecnol\'ogico (CNPq), and also acknowledges financial support from Funda\c{c}\~ao de Amparo \`a Ci\^encia e Tecnologia
do Estado de Pernambuco (FACEPE) under Project No. APQ-0464-1.05/15. 
AA has a fellowship from Conselho Nacional de Desenvolvimento Cient\'{\i}fico e Tecnol\'ogico (CNPq).






\bibliography{references}

\end{document}